\documentclass[journal]{IEEEtran}

\usepackage{graphicx}
\usepackage{epstopdf}
\usepackage{amsmath}
\usepackage{amssymb}
\usepackage{amsthm }
\usepackage{epsfig}
\usepackage{times}
\usepackage{setspace}
\usepackage{bm}
\usepackage{siunitx}
\usepackage{threeparttable}
\usepackage{balance}
\usepackage{tabularx}
\usepackage{subcaption}
\usepackage{xcolor}

\usepackage{array}
\usepackage{fancyhdr}

\usepackage[T1]{fontenc}
\usepackage{times}
\usepackage[mathscr]{eucal}
\usepackage{wrapfig}
\usepackage{multirow}
\usepackage{tablefootnote}
\usepackage{subcaption}

\newcommand{\equals}{\!=\!}
\newcommand{\lteq}{\!\le\!}
\newcommand{\minus}{\!-\!}

\newcommand{\gthan}{\!>\!}
\newcommand{\lthan}{\!<\!}
\newcommand{\plus}{\!+\!}

\newcommand{\SF}{\mathrm{SF}}

\def\BibTeX{{\rm B\kern-.05em{\sc i\kern-.025em b}\kern-.08em
    T\kern-.1667em\lower.7ex\hbox{E}\kern-.125emX}}

\DeclareRobustCommand{\bigO}{%
	\text{\usefont{OMS}{cmsy}{m}{n}O}%
}
\usepackage[ruled,vlined]{algorithm2e}
\DeclareMathOperator*{\argmax}{argmax}

\begin{document}


\title{Making the Most of Sporadic Feedback:\\ Low-Complexity Application-Layer Coding for Data Recovery in the Internet of Things}

\author{

    \IEEEauthorblockN{Vatsalya Chaubey and Siddhartha S. Borkotoky, \textit{Member, IEEE}}
    
    \thanks{This work was supported by the Science
    and Engineering Research Board (SERB), Government of India, through its
    Start-up Research Grant (SRG) under SRG/2020/001491. 
    The authors are with  the Indian Institute of Technology Bhubaneswar, Khordha 752050, India. (Email: vc13@iitbbs.ac.in, borkotoky@iitbbs.ac.in)
    }

}

\maketitle
\thispagestyle{fancy}

\begin{abstract}
We propose application-layer coding schemes to recover lost data in delay-sensitive uplink (sensor-to-gateway) communications in the Internet of Things. Built on an approach that combines   retransmissions and forward erasure correction, the proposed schemes' salient features include low computational complexity and the ability to exploit sporadic receiver feedback for efficient data recovery. Reduced complexity is achieved by keeping the number of coded transmissions as low as possible and by devising a mechanism to  compute the optimal degree of a coded packet in $\bigO(1)$. Our major contributions are: (a) An enhancement to an existing scheme called \textit{windowed coding}, whose complexity is greatly reduced and data recovery performance is improved by our proposed approach. (b) A technique  that combines elements of windowed coding with a new feedback structure to further reduce the coding complexity and improve data recovery. (c) A coded forwarding  scheme in which a relay node provides further resilience against packet loss by overhearing source-to-destination communications and making forwarding decisions based on overheard information.        
\end{abstract}


\section{Introduction}
In Internet of Things (IoT) applications spanning a variety of domains -- such as smart city~\cite{MOC20}, smart agriculture~\cite{CFI19}, industry 4.0~\cite{SGI21}, to name a few -- a common requirement is to wirelessly transfer data to a gateway from  low-cost, battery-powered sensor nodes with modest computational capabilities. In industrial IoT, for example, monitoring and predictive maintenance applications  require sensors installed in manufacturing, transportation, and storage units to measure physical parameters and transmit their readings to a gateway, which forwards the data for further processing, analysis, and decision making~\cite{CBZ20}.        

With the inevitable packet losses that occur over wireless links, transferring sensor data to the gateway with an acceptable probability of message delivery is a challenge~\cite{CBZ20}. Conventional ARQ-type reliability mechanisms employ receiver feedback for retransmission of undelivered messages. However, obtaining regular feedback in IoT settings may be infeasible due to two reasons. First, IoT communications in the unlicensed bands are subject to strict duty-cycle limitations~\cite{RKS17}, and acknowledging every uplink message without exceeding  duty-cycle limits at the gateway becomes infeasible as the number of sensors grows. Furthermore, when the gateway is half-duplex -- which is the case for almost all commercial off-the-shelf units --  frequent feedback transmissions may in fact worsen delivery performance in dense IoT deployments~\cite{PRK17}. Another  reliability-enhancing mechanism is forward-erasure correction, in which the source proactively transmits redundant information to counter packet losses. However, duty-cycle and energy constraints at the sensors limit the amount of redundancy that may be sent. Besides, redundancy computation adds some complexity at the sensors. Another challenge is that many monitoring use cases are delay sensitive, that is, a message \textit{expires} (ceases to be of interest to the recipient) after some time. Any retransmission or redundancy transmission involving the message must occur before its expiry.              

In this paper, we provide methods to improve the probability of sensor data delivery in the presence of irregular receiver feedback by combining retransmissions and forward-erasure correction via \textit{application-layer coding}~\cite{SaR19}. Our design philosophy is guided by a desire to have low computational complexity at the sensors, to make the most of the information gleaned from the occasional feedback messages, and to operate within the constraints imposed by the application's delay sensitivity and the regulatory limits on transmitter duty-cycle.

\subsection{Application-Layer Coding}
In application-layer coding, the source transmits redundancy in the form of \textit{coded symbols}, which are linear combinations of the \textit{information symbols}. (In our context, an information symbol is the binary representation of a sensor measurement.) At the receiver, recovering  information symbols from coded symbols is equivalent to solving linear equations~\cite{SaR19}. The number of information symbols combined to form a coded symbol is called the latter's \textit{degree}. Depending on the coding scheme, the degree may be chosen in a random or deterministic fashion, and the linear combinations may be computed in Galois Fields (GFs) of different orders.

The best known example of application-layer codes is the family of fountain codes~\cite{Mac05} -- such as Luby Transform (LT) codes~\cite{Lub02} and Raptor codes~\cite{Sho06} -- which have been used in a variety of domains including multimedia broadcasting~\cite{3GPP}, distributed storage~\cite{DPR06}, big data processing~\cite{MCS20}, and blockchains~\cite{YGA22}. To transfer $K$ information symbols, the fountain encoder transmits a potentially infinite number of coded symbols, each being the bitwise XOR of $D$ randomly chosen information symbols, where $D$ is also random with a certain degree distribution (e.g., the robust soliton distribution for LT codes) and varies from one coded symbol to the next. Once the receiver accumulates  a set of coded symbols that includes $K$  linearly independent combinations of information symbols, techniques such as Gaussian elimination, belief propagation~\cite{Lub02}, or a combination thereof~\cite{Sho06} are used to retrieve the information symbols.

Despite their strong erasure-correction capability, fountain codes are not well-suited for  uplink IoT communications in which the information symbols (e.g., sensor measurements) are generated in real time and their utility expires after certain duration (as a result, a large pool of $K$ information symbols is not available at all instants). Therefore, for this work, the semi-random approaches introduced in~\cite{BSR20} are of particular interest. In~\cite{BSR20}, information from sporadic receiver feedback is used to make retransmission and coding decisions, to judiciously choose the degree of coded symbols, and to pick the information symbols to be combined in GF(2) from a continuously changing pool. This approach provides better recovery performance and lower complexity (in terms of average number of XORs performed per transmission) compared to blind  schemes in which end devices transmit coded symbols at every opportunity. 

\pagestyle{plain}
\subsection{Contributions of This Paper}
We utilize elements of the \textit{windowed coding} (WC)  approach of~\cite{BSR20} to propose data recovery schemes aimed at delay-sensitive IoT applications with limited receiver feedback. Our  novel contributions include the derivation of a computationally lightweight degree-selection procedure, proposal of a new feedback structure to devise  schemes with lower complexity and higher reliability, and the development of a  relay-aided application-layer coding mechanism for further reliability improvements. Specifically, we propose the following techniques:
\begin{enumerate} 
    \item \textit{Improved Windowed Coding} (IWC): An improved version (lower complexity, higher data recovery) of WC in which: (a) the complexity of degree selection is reduced from approximately $\bigO(\delta^\delta)$ to $\bigO(1)$ -- where $\delta$ is the number of unexpired symbols at a given instant -- thus greatly reducing memory and computational requirements at the source, and (b) the coding degree employed in the absence of feedback is modified to improve the data recovery rate.   
    
    \item \textit{IWC with modified feedback} (IWC-MF): An application-layer coding scheme with modifications to the feedback structure of IWC so that more information is conveyed using the same number of feedback bits. The modification reduces the number of coded symbol transmissions, improves the data-recovery performance, and reduces the computational load.
    
    \item \textit{Relay-aided IWC} (IWC-R): A mechanism in which a forwarding node (relay) overhears the communications between the source and the destination, and forwards uncoded information symbols or constructs and transmits new coded symbols. We demonstrate that this approach provide higher data recovery than a system without relays or with a conventional decode-and-forward relay that that does not perform coding. 
\end{enumerate}

\section{Related Works}
A scheme suitable for application-layer coding on real-time data streams is proposed in~\cite{DKF13} under the assumption that the source receives feedback for every transmission. An application-layer coding scheme for delay-sensitive multimedia traffic is given in~~\cite{FEL19}. The scheme operates in GF(256), which is computationally burdensome for unsophisticated end devices with limited memory and compute resources. In~\cite{MKR20}, the authors propose an application-layer coding scheme in GF(2) for LoRa IoT applications, assuming complete absence of feedback. Retransmissions and coded transmissions are combined without using receiver feedback in~\cite{SHS20}. The WC scheme and its variant called selective coding (SC) of~\cite{BSR20} exploit occasionally available receiver feedback to combine retransmissions and coding in GF(2).

Compared to single-hop communications, there are  fewer works on application-layer coding in relay-aided IoT settings. In~\cite{TBK20}, multiple forwarding nodes in a LoRa network employ random linear network coding to forward messages from end devices to a gateway. Significant coordination among the relays is required, and the availability of regular feedback is assumed. A coded-relaying scheme in which a relay forwards XOR sums of overheard sensor messages without utilizing any feedback is proposed in~\cite{BAC21}.

\section{System Model}
\label{sec:system_model}

\begin{figure}
    \centering
    \includegraphics[scale=0.24, bb=140 180 820 480]{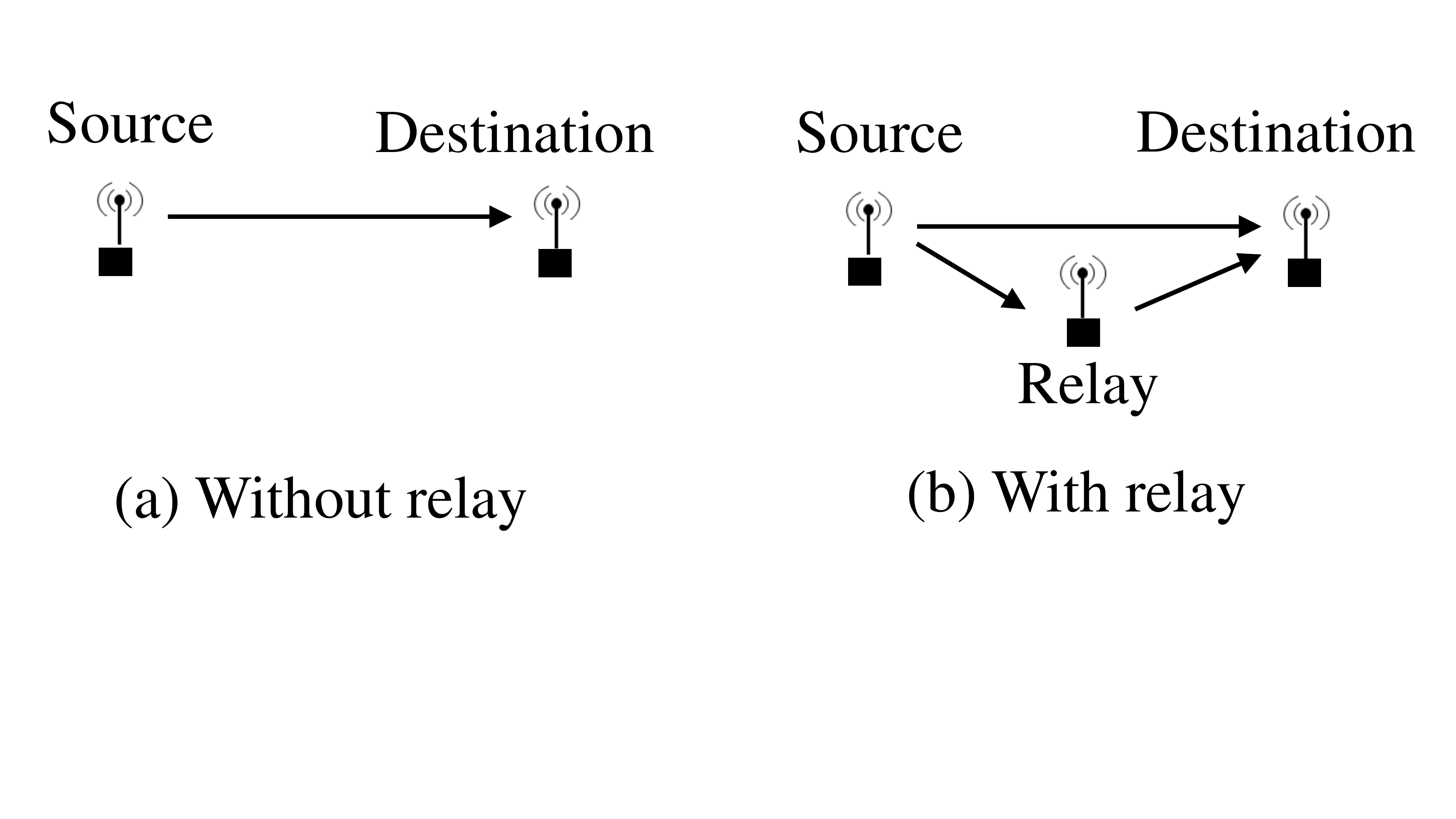}
    \caption{Packet communications with and without a relay.}
    \label{fig:NetDiag}
\end{figure}

\begin{figure}
    \centering
    \includegraphics[scale=0.26, bb=140 280 820 520]{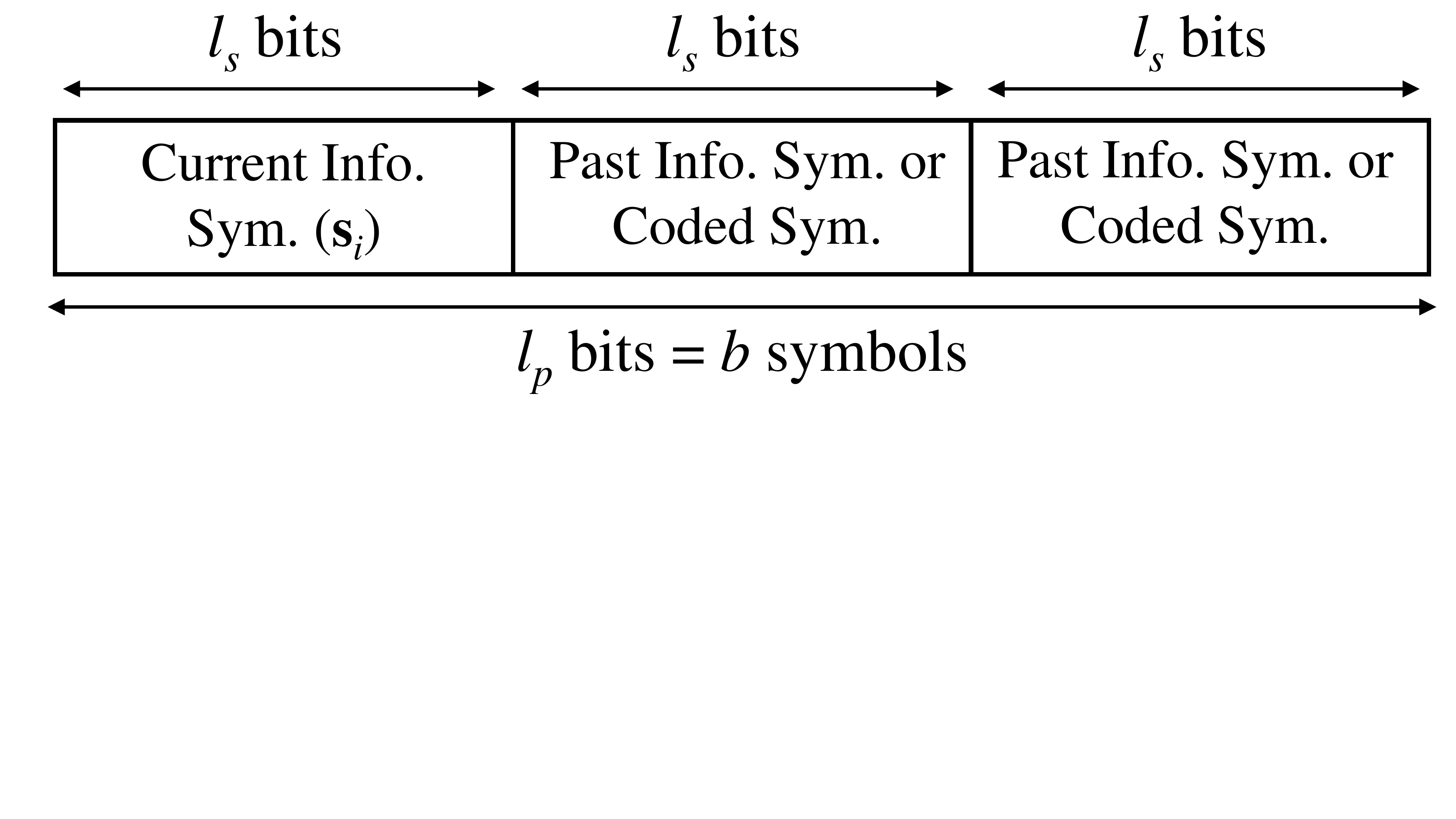}
    \caption{Packet payload structure for WC.}
    \label{fig:packet_structure_WC}
\end{figure}

We consider unicast networks comprising a \textit{source}, a \textit{destination},  and occasionally a \textit{relay  node}, as shown in  Fig.~\ref{fig:NetDiag}. The source generates a continuous stream of information symbols (e.g., sensor measurements), each $l_s$-bits long. The information symbol $\mathbf{s}_i$ generated at time instant $i$ triggers the transmission of a packet $\mathbf{p}_i$ that contains $\mathbf{s}_i$ in the payload. The packet may also contain past information symbols and coded symbols, which are XORs of past information symbols chosen at random from a set of symbols called the \textit{coding set}. Each coded symbol is also $l_s$-bits long. The maximum number of symbols that can fit in a packet is $b \equals l_p/l_s$, where $l_p$ is the maximum possible payload size (in bits), which depends on the lower-layer communication protocols, duty-cycle regulations, and transmitter energy constraints. The packet  structure is illustrated in Fig.~\ref{fig:packet_structure_WC}.

\begin{table} [t]
\caption{Summary of key notation}
\begin{tabular}{ c | c  }
    \hline
    Notation & Description \\ \hline
    $\mathrm{s}_i$ &  Information symbol no. $i$\\ \hline
    $\mathbf{p}_i$ &  Packet no. $i$\\ \hline
    $b$ & Maximum number of symbols \\
    & a packet can carry \\ \hline
    $u$ &  Sequence number of oldest unexpired \\
     & and undelivered information symbol ($\mathrm{s}_u$)\\  \hline
    $\beta$ & Number of unexpired information symbols\\
    & not yet delivered \\ \hline
    $\delta$ & Delay tolerance \\ \hline
    $d_{\mathrm{nf}}$ &  No-feedback degree \\  \hline
    $R_m$ & Max. no. of symbols relay can store \\ \hline
    $R_t$ & Relay threshold \\ \hline
    $p_\mathrm{fb}$ & Feedback-reception probability \\ \hline
\end{tabular}
\label{Notation}
\end{table}

We consider random erasure channels in which a packet is either received correctly or is erased (lost) in a random fashion.  Following the successful reception of a packet, the destination stores the uncoded information symbols contained therein in its memory, and uses the coded symbols to recover previously undelivered information symbols. We say that an information symbol is \textit{delivered} if its uncoded version is in a received packet or if it was recovered from a coded symbol. Specifically, a coded symbol of degree $d$ results in the recovery of an information symbol if exactly $d \minus 1$ of the XORed information symbols were already delivered. A coded symbol that does not result in the recovery of an information symbol is immediately discarded. (While it is possible to buffer such coded symbols at the destination for future decoding attempts, it increases memory requirements -- an issue that is exacerbated as the gateway needs to handle more and more end devices -- thus  posing a scalability problem.)

As in~\cite{BSR20}, the choice to send multiple symbols in the same packet is motivated by a desire to reduce the transmission overhead. Since most sensor data are small (a few bytes), the packet overhead (synchronization preamble and various protocol headers) can be substantial relative to the payload size. Including multiple symbols in one packet leads to much lower energy consumption and smaller duty cycles than sending each symbol in a separate packet. For example, transmitting a 3-byte information symbol and a 3-byte coded symbol in two separate packets using the LoRa physical layer~\cite{Sem13} results in 67\,\% higher duty cycle and energy expenditure compared to sending them in the same packet. In addition, it may be possible to include redundancy in a packet without increasing the packet length. For example, a LoRa packet's duration for spreading factor 10 is constant for all payload sizes between 1 byte and 4 bytes~\cite{Sem13}. Thus, up to three coded symbols can be included with a 1-byte information symbol without impacting the packet length, duty cycle, and energy expenditure.

We denote by $\delta$  the \textit{delay tolerance} of the application. At time instant $i$, the information symbols generated prior to time instant $i \minus \delta$ are treated as expired.

After a transmission, a feedback from the destination may arrive at the sender. We assume a cumulative feedback structure that carries the sequence number of the oldest undelivered but unexpired packet. Feedback arrival is modeled as a stochastic process, characterized by the  \textit{feedback reception probability}, denoted by $p_\mathrm{fb}$. The feedback-reception probability is a product of two terms,  the probability that the receiver is able to transmit a feedback packet and the probability that a feedback packet, if transmitted, arrives correctly at the sender.

\section{Review of Windowed Coding (WC)}
\label{sec:WC}

\begin{algorithm}[b]
	\SetAlgoLined
	\KwResult{Message delivery at the receiver}
	Generate information symbols at the sender\;
	\For{all information symbols at the sender}{
	    include current information symbol in packet\;
		\uIf{feedback was received for previous packet}
		{
		    include oldest undelivered symbol in packet\;
            \uIf{all missing symbols fit in the packet} {
            include uncoded symbols\;
            }
            \uElseIf{all symbols that fit are guaranteed to be missing symbols}
            {
            include uncoded symbols\;
            }
            \Else{
            include coded symbols with degree given by~\eqref{eq:original-opt-d} \;
            }
			
		}
		\Else
		{
			include coded symbols having  random degrees;
		}
	}
	\caption{Windowed Coding (WC)}
	\label{algo:WC}
\end{algorithm}

We provide a brief overview of WC here.
For a detailed description, the reader is referred to Section IV.A of~\cite{BSR20}.

\begin{figure}
    \centering
    \includegraphics[scale=0.26, bb=140 300 820 520]{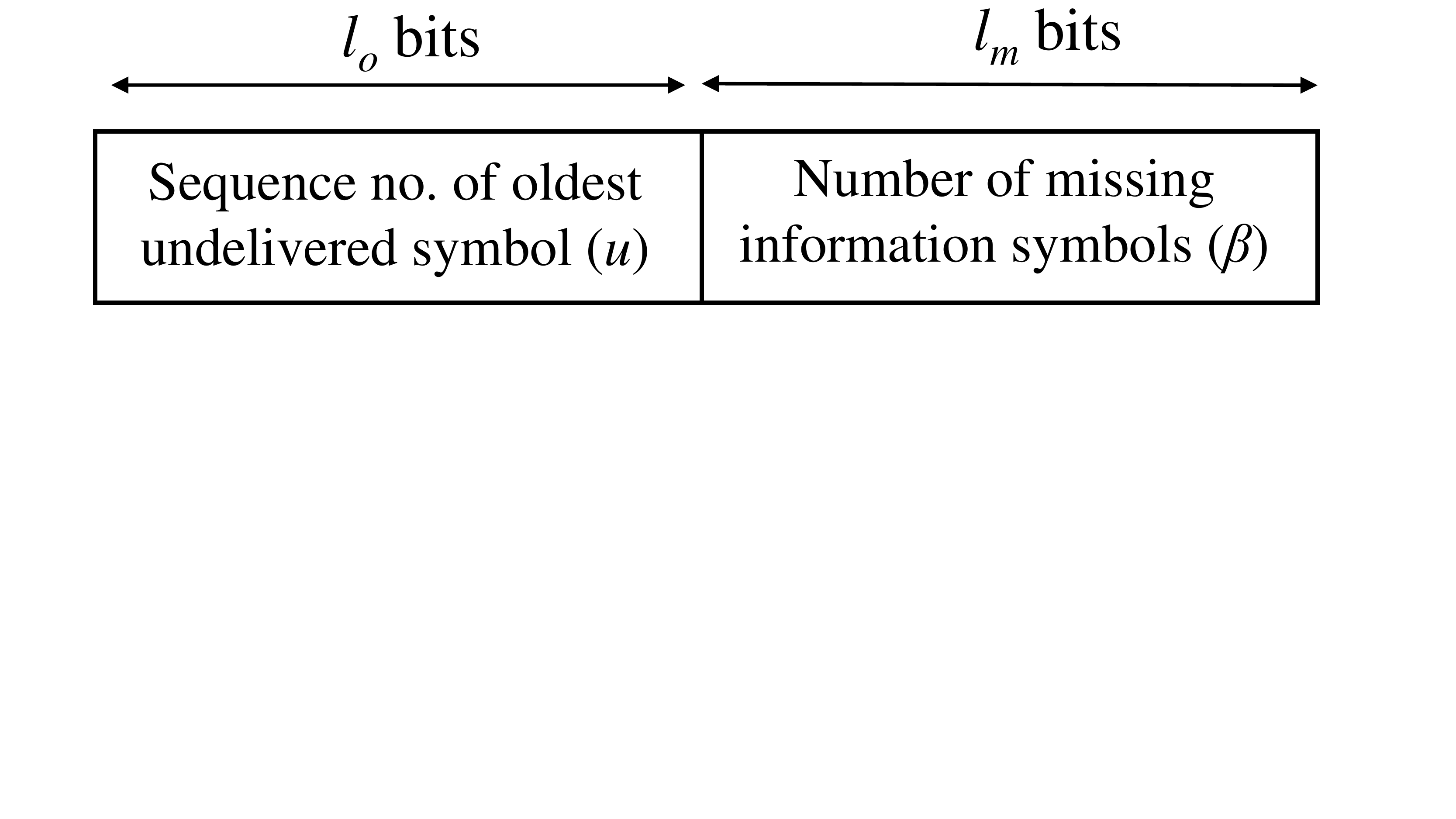}
    \caption{Feedback packet structure for WC.}
    \label{fig:feedback_WC}
\end{figure}

\begin{figure}
    \centering
    \includegraphics[scale=0.24, bb=140 100 820 520]{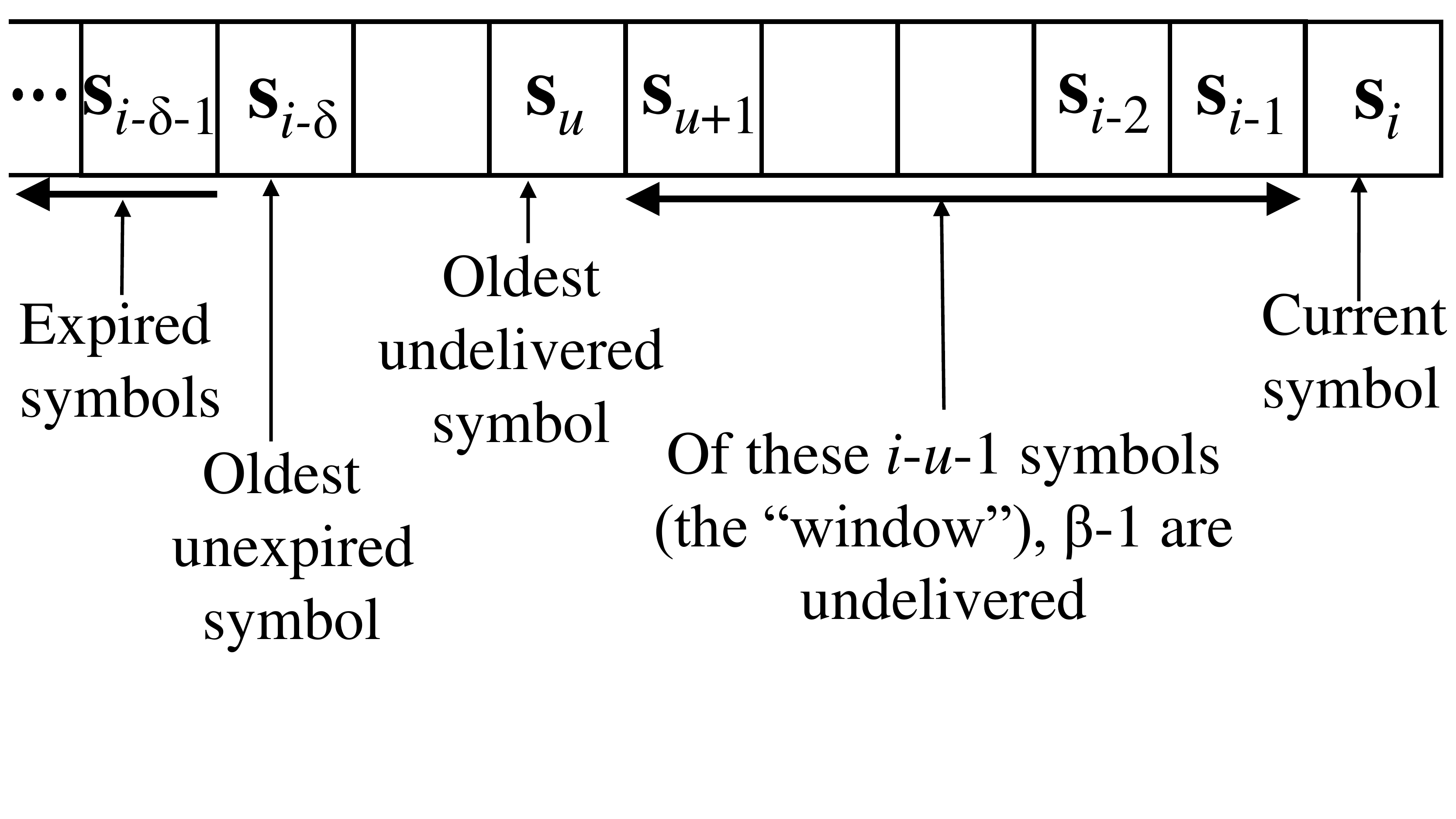}
    \caption{Sender's knowledge following feedback reception.}
    \label{fig:delivery_status_sender}
\end{figure}

First consider WC's feedback structure, as shown in Fig.~\ref{fig:feedback_WC}. A feedback packet contains the sequence number $u$ of the oldest unexpired information symbol that is not yet delivered, and the total number $\beta$ of undelivered and unexpired information symbols at the instant the feedback is sent. The sender's view of the symbol delivery status at the time of transmitting $\mathbf{p}_i$ is shown in Fig.~\ref{fig:delivery_status_sender}, assuming a feedback was received after  $\mathbf{p}_{i-1}$. The sender knows that $\mathbf{s}_u$ is the oldest undelivered symbol, and it knows that over the ``window'' spanning  $\mathbf{s}_{u+1}$ through $\mathbf{s}_{i-1}$, the receiver is missing exactly $\beta \minus 1$ symbols (but it does not know which ones, unless $\beta \equals i \minus u$, in which case all symbols in the window are missing.)  

In source packet $\mathbf{p}_i$ (see Fig.~\ref{fig:packet_structure_WC}), the first symbol is the current information symbol $\mathbf{s}_i$. The remaining $b \minus 1$ symbols depend on whether a feedback was received following $\mathbf{p}_{i-1}$.

Suppose a feedback was received following $\mathbf{p}_{i-1}$ and as per the feedback, $u \equals i$. Then $\mathbf{p}_i$ contains only $\mathbf{s}_i$. If $u \neq i$, then $\mathbf{p}_i$ contains both $\mathbf{s}_i$ and $\mathbf{s}_u$. Now if $\beta \equals 1$, then there are no missing symbols in the window, and nothing is included in the remaining $b \minus 2$ symbol spaces in the packet. If $\beta \gthan 1$ and $i \minus u \minus 1 \lteq b \minus 2$, all symbols in the window can be included in the packet without regard to which ones are missing. If $i \minus u \minus 1 \gthan b \minus 2$ and $\beta \equals i \minus u$, all symbols in the window are missing but not all of them will fit in the packet;  the oldest $b \minus 2$ symbols are included in this case. Note that all the above involve the transmission of uncoded information symbols.

By contrast, coded symbols are transmitted if $i \minus u \gthan b \minus 1$ and $1 \lthan \beta \lthan i \minus u$. This corresponds to a situation in which $\beta \minus 1$ out of the $i \minus u \minus 1$ symbols in the window are missing, but there is not enough space in the packet to send all $i \minus u \minus 1$ uncoded symbols. Instead, $b \minus 2$ coded symbols of degree $d$ are produced from the coding set $\{\mathbf{s}_{u+1}, \mathbf{s}_{u+2},\ldots, \mathbf{s}_{i-1}\}$, where             
\begin{equation}
	d = \argmax_{d'} \frac{\binom{\beta - 1}{1}\cdot \binom{i-u-\beta}{d'-1}}{\binom{i-u}{d'}}.
	\label{eq:original-opt-d}
\end{equation}
The objective of~\eqref{eq:original-opt-d} is to maximize the probability that exactly one of the $d$ XORed information symbols belongs to the set of $\beta - 1$ undelivered symbols in the coding set. This is equivalent to maximizing the probability that the coded symbol allows the recovery of one information symbol.

If no feedback arrives  following $\mathbf{p}_{i-1}$, then there are two possibilities. Let $u_0$ be the sequence number of the oldest unexpired symbol at time $i$ and let $u_l$ be the sequence number of the oldest undelivered symbol as per the most recent feedback. Now define $u' \equals \max\{u_0, u_l\}$. If the set $\mathbf{w} = \{\mathbf{s}_{u'}, \mathbf{s}_{u'+1},\ldots, \mathbf{s}_{i-1}\}$ contains $b \minus 1$ or fewer elements, then $\mathbf{p}_i$ contains $\mathbf{s}_i$ followed by each of the elements of $\mathbf{w}$. Else, it contains $\mathbf{s}_i$ followed by $b -1$ coded symbols produced from the coding set $\mathbf{w}$, with degrees chosen uniformly at random from the set of integers $\{1,2,\ldots,i \minus u'\}$. The basic steps of WC are listed in Algorithm~\ref{algo:WC}.

\section{Improved Windowed Coding (IWC)}
\label{sec:IWC}

\begin{algorithm}[t]
	\SetAlgoLined
	\KwResult{Message delivery at the receiver}
	Generate information symbols at the sender\;
	\For{all information symbols at the sender}{
	    include current information symbol in packet\;
		\uIf{feedback was received for previous packet}
		{
		    include oldest undelivered symbol in packet\;
            \uIf{all missing symbols fit in the packet} {
            include uncoded symbols\;
            }
            \uElseIf{all symbols that fit are guaranteed to be missing symbols}
            {
            include uncoded symbols\;
            }
            \Else{
            include coded symbols with degree given by~\eqref{eq:degree_oneshot} \;
            }
			
		}
		\Else
		{
			include coded symbols having  degree $d_{\mathrm{nf}}$;
		}
	}
	\caption{Improved Windowed Coding (IWC)}
	\label{algo:IWC}
\end{algorithm}

We now describe some shortcomings in WC and present remedial measures, which result in IWC. The IWC scheme uses the same feedback  structure  as WC (see  Fig.~\ref{fig:feedback_WC}). The primary changes are in the degree-selection procedure.  

\subsection{Degree Selection Following Feedback Reception}

Degree  selection via~\eqref{eq:original-opt-d} is a computationally intensive task. With increasing delay tolerance $\delta$, the number of $\binom{n}{k}$ terms to be evaluated also increases. The complexity of one such $n \choose k$ operation is approximately $\bigO(\min\{n^k, n^{n-k}\}) \simeq \bigO(n^k)$, so the complexity of degree-selection for a packet is
\[
\bigO\left(\sum \limits_{d'=1}^{d'=i-u-\beta +1}(i-u-\beta)^{d'-1} \cdot (i-u)^{d'}\right) \simeq \bigO(\delta ^ {\delta}),
\]
which may be excessive for memory constrained and battery power end devices. Even if the optimal degrees are precomputed and stored in memory, the size of the required look-up table is proportional to $\bigO(\delta^2)$. 

For IWC, we propose a technique to compute the optimal degree without requiring any $n \choose k$ operations and without searching over all possible candidate degrees. Specifically, we propose to use the following degree for a coded symbol
\begin{equation} \label{eq:degree_oneshot}
	d = \min\left\{\left\lfloor{\frac{i-u}{\beta - 1}}\right\rfloor, i-u-\beta\right\}.
\end{equation}
As proved in the Appendix, degree selection via~\eqref{eq:degree_oneshot} and~\eqref{eq:original-opt-d} are equivalent. However, \eqref{eq:degree_oneshot} 
can be computed in a single shot without the need for any $n \choose k$ operations, thereby reducing the overall complexity and memory requirement to $\bigO(1)$, as opposed to the computationally demanding evaluation of~\eqref{eq:original-opt-d}.

\vspace{-2mm}
\subsection{Degree Selection in the Absence of Feedback}

Recall that in the absence of feedback following a transmission, WC first checks if it needs to code for the next packet. If so, coded symbols with randomly chosen degrees are produced. However, as we demonstrate in Section~\ref{sec:perf_res},  such random selection does not always provide good performance.

Let the \textit{no-feedback degree} ($d_{\mathrm{nf}}$) denote the degree used for the coded symbols in a packet that is sent following a transmission for which no feedback was received.  
Unlike the case in which feedback is received, determining the optimal degree in the  absence of feedback is an intractable problem to the best of our knowledge. From extensive simulations, we have found $2 \lteq d_{\mathrm{nf}} \lteq 4$  to provide near-optimal performance over a wide range of scenarios. For our purposes, we use \mbox{$d_{\mathrm{nf}} \equals 2$}. Note that a lower-degree coded symbol requires fewer XOR operations, motivating our choice of $d_{\mathrm{nf}} \equals 2$.

\section{IWC With Modified Feedback (IWC-MF)}
\label{sec:IWC-MF}

\begin{figure}
    \centering
    \includegraphics[scale=0.26, bb=140 200 820 400]{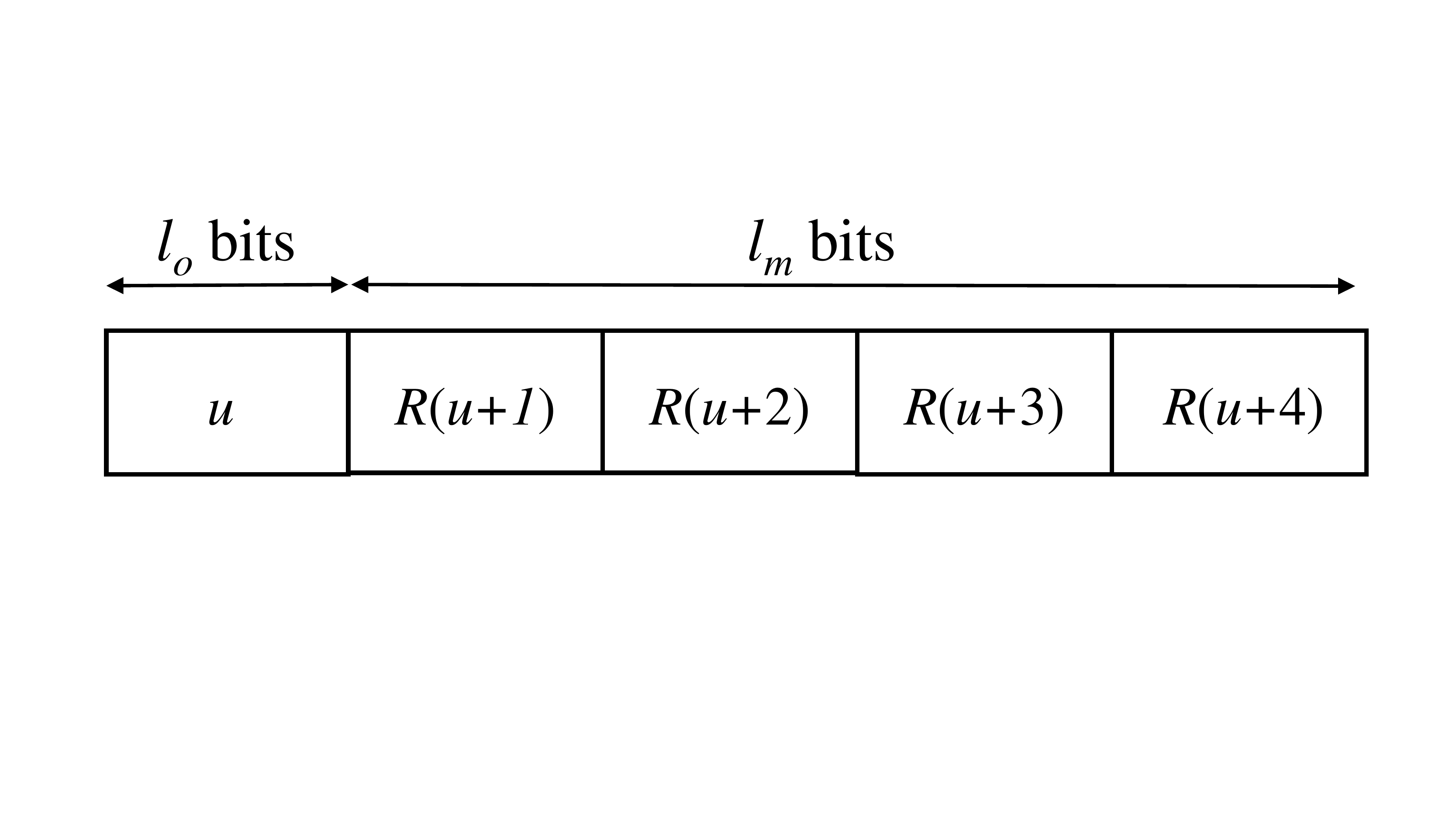}
    \caption{Feedback  structure for IWC-MF (shown for $l_m \equals 4)$.}
    \label{fig:feedback_IWC-MF}
\end{figure}

IWC-MF is derived by changing the feedback structure of WC and IWC. Recall that the feedback in WC has two main components~-- the sequence number of the oldest undelivered-but-unexpired packet and the number of missing symbols. Let the number of bits reserved for these two fields be $l_o$ and $l_m$, respectively. The $l_m$ bits convey the overall state of the receiver in terms of the number of missing symbols, but they do not specify which symbols are missing.

We now define the new feedback structure of Fig.~\ref{fig:feedback_IWC-MF} with the same number of total bits as in IWC. As before, $l_o$ bits are  used for the sequence number $u$ of the oldest undelivered symbol. But the remaining $l_m$ bits, which in IWC carry the number of missing symbols, are now used to convey the delivery status of the symbols $s_{u+1}$ through $s_{u+l_m}$. The feedback contains the binary values of $R(j)$ for $u \plus 1 \lteq j \lteq u \plus l_m$, where $R(j)$ is 0 if $\mathbf{s}_j$ is yet undelivered, and 1 otherwise. Upon receipt of the feedback, the sender includes the missing symbols (or as many missing symbols as there is space) in the next packet. Ideally, we would want $l_m \geq \delta$, so that information about the state of every unexpired symbol is known. But even with $l_m < \delta$, we obtain a performance improvement over IWC since we  transmit the uncoded versions of some of the missing symbols. Note that we prioritize the transmission of the older symbols, which are due to expire earlier. In contrast, when we did not know the exact state and resorted to coding, the symbols are chosen at random irrespective of their order, which can lead to older missing symbols not being chosen and thus expiring, while some other symbols were sent. 

When a feedback does not arrive, IWC-MF resorts to the IWC philosophy, that is, it transmits coded symbols with degree $d_{\mathrm{nf}}$. Thus,  \mbox{IWC-MF} avoids coding when feedback is received and transmits coded symbols only in the absence of feedback, leading to fewer computational operations at the source. The steps are enumerated in Algorithm~\ref{algo:IWC-MF}.

\begin{algorithm}[t]
	\SetAlgoLined
	\KwResult{Message delivery at the receiver}
	Generate information symbols at the sender\;
	\For{all information symbols at the sender}{
	    include current information symbol in packet\;
		\uIf{feedback was received for previous packet}
		{
            Create a packet with uncoded missing information symbols as learnt from feedback\;
		}
		\Else
		{
			include coded symbols having  degree $d_{\mathrm{nf}}$;
		}
	}
	\caption{Improved Windowed Coding With\\ Modified Feedback  (IWC-MF)}
	\label{algo:IWC-MF}
\end{algorithm}

\section{Relay-Aided IWC (IWC-R)}
The IWC-R scheme  improves the data recovery performance of IWC by introducing a relay node  as shown in network (b) in Fig.~\ref{fig:NetDiag}. The source  is not necessarily aware of the relay's presence, and the destination does not send any explicit feedback to the relay. The relay attempts to overhear the communications between the source and the destination, and makes forwarding decisions based on the information extracted from the overheard packets.  

\subsection{Overview of Relay Operation}
The relay continually alternates between two phases, the \textit{reception phase} in which it tries to overhear any communication between the source and the destination, and the \textit{forwarding phase} in which it forwards source messages to the destination.  

During the reception phase, the relay buffers the uncoded information symbols contained in the overheard source packets. Coded symbols, if any, are discarded. From overheard feedback packets, the relay notes the oldest undelivered sequence number $u$. Note that a source packet carries at least one uncoded information symbol and a feedback contains $u$ regardless of whether WC, IWC, or IWC-MF is used on the source--destination link. Thus, the relay operation is compatible with all three. In our illustrations, we employ IWC.

Once the relay buffers $R_t$ new information symbols since its last transmission, it enters the forwarding phase and transmits \textit{one} symbol, decided in a manner described in the following subsection. Thus, one symbol is transmitted for every $R_t$ received symbol. We call $R_t$ the  \textit{relay threshold}, which can be tuned to control the relay's duty cycle. After the transmission, the reception phase resumes and the cycle continues.   

\subsection{Transmitted Symbol by Relay}
Let $\mathbf{p}_j$ be the last packet overheard during a certain reception phase. For the subsequent forwarding phase, the relay first checks if it overheard the destination's feedback following $\mathbf{p}_{j-1}$. If so, and if the oldest undelivered symbol $\mathbf{s}_{u}$ is in the relay's buffer, the relay forwards $\mathbf{s}_{u}$. 

If no feedback was overheard, or if $\mathbf{s}_{u}$ is older than the oldest packet in the relay's memory, the relay transmits a coded symbol with degree of $\min \{d_{\mathrm{nf}} , m\}$, where $m$ is the number of buffered symbols at the relay.

 \begin{algorithm} [t]
	\label{relay-coding}
	\SetAlgoLined
	\KwResult{Message delivery at the receiver}
	Perform IWC at the source\;
	Store overheard information symbols in relay memory, discard old messages that do not fit in memory\;
	\If{number of symbols stored since last transmission exceeds relay threshold}{
		\eIf{previous feedback is received \textbf{and} oldest undelivered message is in memory}{
			transmit the oldest undelivered symbol\;
		}{
			transmit a coded symbol\;
		}
	}
	\caption{Relay-Aided IWC (IWC-R)}
\end{algorithm}

\section{Performance Metric, Benchmarks, and Simulation Setup}

Our reliability measure is the \textit{delivery failure rate} (DFR),  defined as the  
fraction of information symbols generated at the source that remain undelivered (i.e, are not delivered prior to their expiry). Thus, if the transmission of $N$ unique information symbols are simulated,  of which $M$ are delivered unexpired to the destination, the DFR is \mbox{$(N-M)/N$}.

We evaluate our proposed methods against the following baseline schemes: 
\begin{enumerate} 
    \item \textit{Repetition Redundancy (RR)}: It utilizes receiver feedback but does not employ coding. Packet $\mathbf{p}_i$  contains the current information symbol $\mathbf{s}_i$. If feedback for $\mathbf{p}_{i-1}$ was received, then $\mathbf{p}_i$ also contains the oldest undelivered symbol $s_u$. The remaining  space  is occupied by up to $b \minus 2$ most recent unacknowledged symbols. In the absence of a feedback, $\mathbf{p}_i$ contains $\mathbf{s}_i$ followed by up to $b \minus 1$ most recent unacknowledged symbols. 
    
    \item \textit{Windowed Coding (WC)}: This is the windowed coding scheme of~\cite{BSR20}.  It differs from IWC in terms of degree-computation following a feedback reception and in the choice of degree following the non-arrival of a feedback.
    
    \item \textit{Uncoded Relaying (UC-R)}: This scheme employs a relay node that immediately forwards every received information symbol without performing any coding, while IWC is performed on the source-to-destination link.
\end{enumerate}

We evaluate our schemes for Bernoulli and Gilbert-Elliott (GE) erasure channels. In the former, packet erasures are independent Bernoulli random variables, with $P_s$ denoting the probability of correct reception. The GE channel exhibits correlated (bursty) packet losses, and is based on a first order, two-state Markov chain, the states being  \textit{good} and \textit{bad}. The channel state remains unchanged throughout the duration of a packet, but may change from one packet to the next according to the transition probabilities of the Markov chain. A packet transmitted while the channel is in the good state is successfully received, and it is lost if transmitted during the bad state. The transition probability from good to bad state is $p_{\mathrm{gb}}$ and from bad to good is $p_{\mathrm{bg}}$. For both Bernoulli and GE uplinks, feedback reception is modeled as an independent Bernoulli event, with $p_{\mathrm{fb}}$ denoting the probability that a feedback is received following a transmission.    

We simulate each scheme for the transmission of $100,000$ symbols. Table~\ref{tab:default_sim_parameters} lists the default parameter values that are used, unless stated otherwise. 

\begin{table}[]    \caption{Default simulation  parameters.}
    \centering
    \begin{tabular}{c|c}
        Parameter & Value \\ \hline
        Delay tolerance ($\delta$) & 16 \\
        Feedback-reception probability ($p_\mathrm{fb}$) & 0.25 \\ 
        Good-to-bad transition prob. for GE channel ($p_\mathrm{gb}$) & 0.25 \\ 
        Max. symbols per source packet ($b$)  & $3$ \\
        No-feedback degree ($d_{\mathrm{nf}}$) & 2 \\
        Max. no. of symbols relay can store ($R_m$) & 16 \\
        Feedback bits to convey missing symbols ($l_m$) & 4
    \end{tabular}
    \label{tab:default_sim_parameters}
\end{table}

\section{Results and Discussions}
\label{sec:perf_res}

\begin{figure}
    \centering
    \includegraphics[scale=0.56, bb=40 270 820 550]{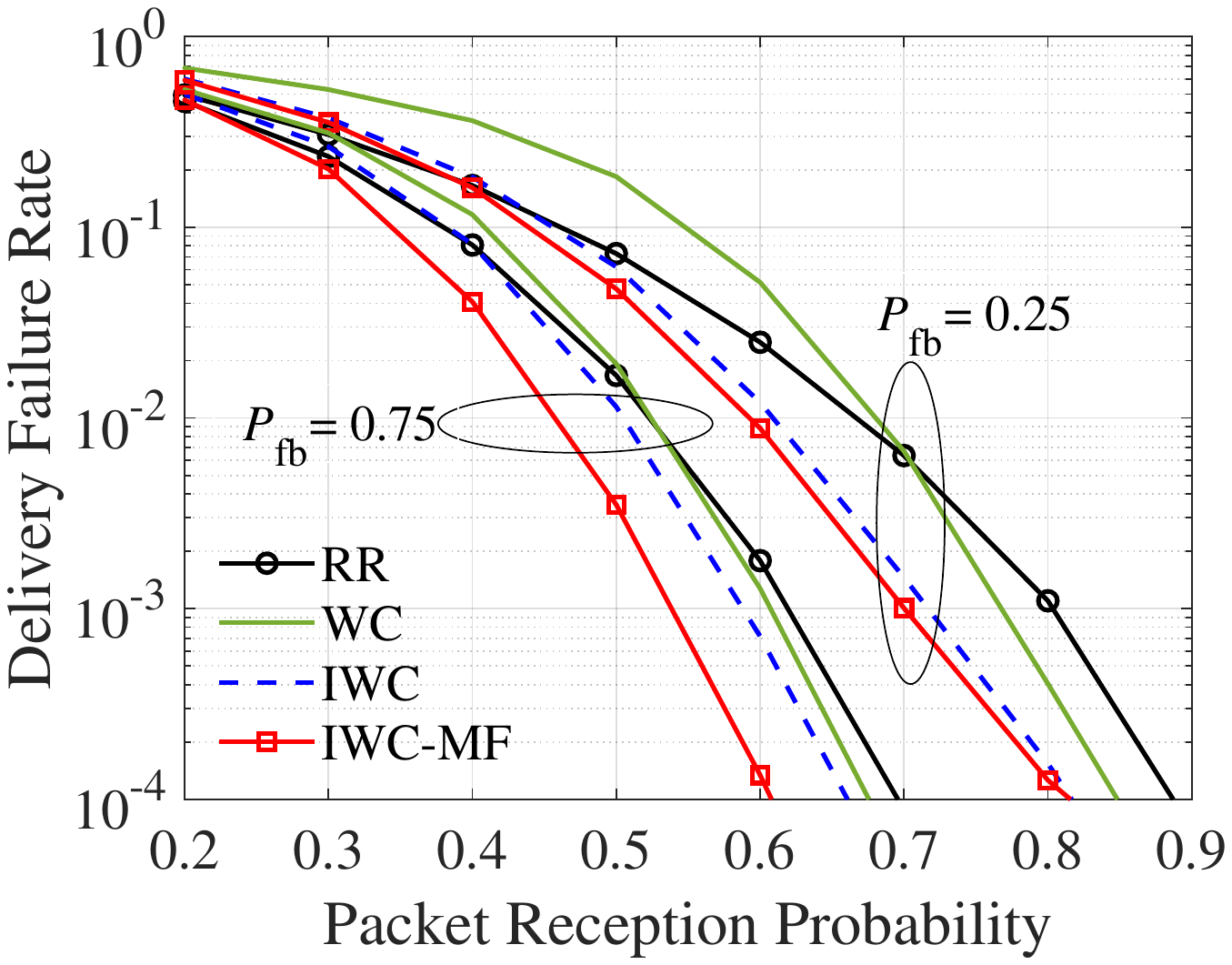}
    \caption{Delivery failure rates on a Bernoulli erasure channel.}
    \label{fig:DFR_NoRelay_Bernoulli}
\end{figure}

Fig.~\ref{fig:DFR_NoRelay_Bernoulli} compares the performance of the proposed  IWC and IWC-MF schemes to that of RR and WC over a Bernoulli erasure channel. Results are shown for feedback-reception probabilities of 0.25 and 0.75. For each scheme, a higher $P_{\mathrm{fb}}$ results in lower delivery failure rate for any given value of the packet success probability on the uplink. For each $P_{\mathrm{fb}}$, we observe that the proposed schemes outperform the baselines over a wide range of packet reception probabilities. We also observe that for $P_{\mathrm{fb}} \!=\! 0.25$,   both IWC and IWC-MF significantly outperform RR and WC. For $P_{\mathrm{fb}} \!=\! 0.75$, IWC-MF provides higher performance gains compared to IWC. 

It is interesting to note that the baseline WC scheme provides performance gains over RR only when the packet reception probability is high. Note that much higher performance gains for WC over RR were reported in~\cite{BSR20}. We found that to be the case indeed when the destination stored old coded symbols (the ones that did not immediately led to the recovery of an information symbol) in its memory for future use. By contrast, our simulation model  assumes that  coded symbols that do not immediately recover an information symbol are discarded. This is done to avoid excessive memory requirements as the network grows, requiring the destination (gateway)  to receive from a large number of sources (sensors).

\begin{figure}
    \centering
    \includegraphics[scale=0.56, bb=40 270 820 550]{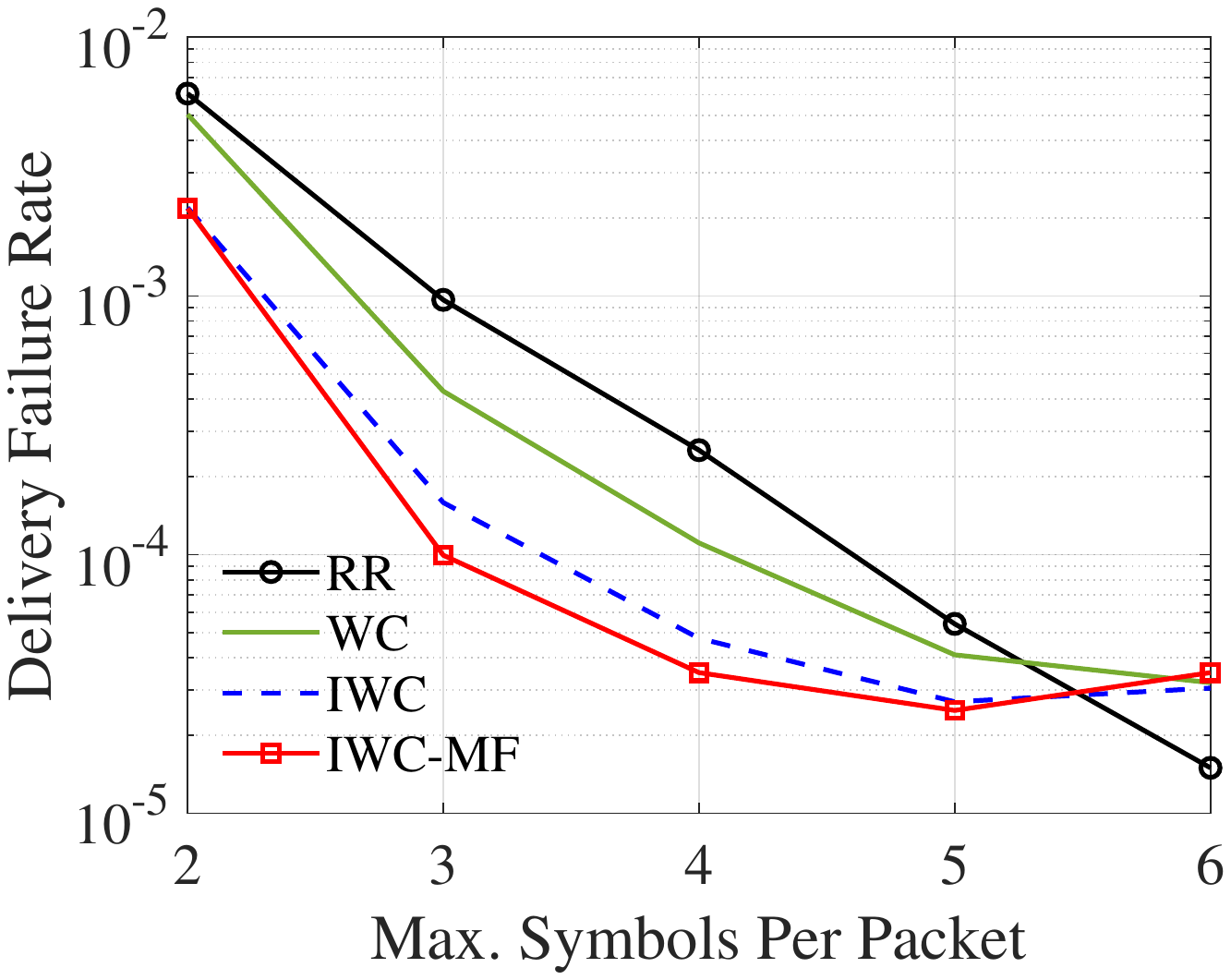}
    \caption{Impact of the maximum number of symbols per packet.}
    \label{fig:DFR_vs_b}
\end{figure}

Fig.~\ref{fig:DFR_vs_b} shows the performance of the schemes as the maximum number of symbols per packet ($b$) is varied. For $b \lteq 5$, all forms of coding provides higher data recovery than RR, and the proposed schemes outperform WC by up to an order of magnitude  for $b \equals 3$ and $b \equals 4$. The performance of all schemes improve with $b$, as would be intuitively expected. But for $b \gthan 5$, the performance of the coding schemes tend to saturate, while that of RR continues to decrease. In practical IoT applications with energy and duty-cycle constrained IoT nodes, we expect smaller payload sizes to be more prevalent, and coding outperforms RR in that regime.

\begin{figure}
    \centering
    \includegraphics[scale=0.56, bb=40 270 820 550]{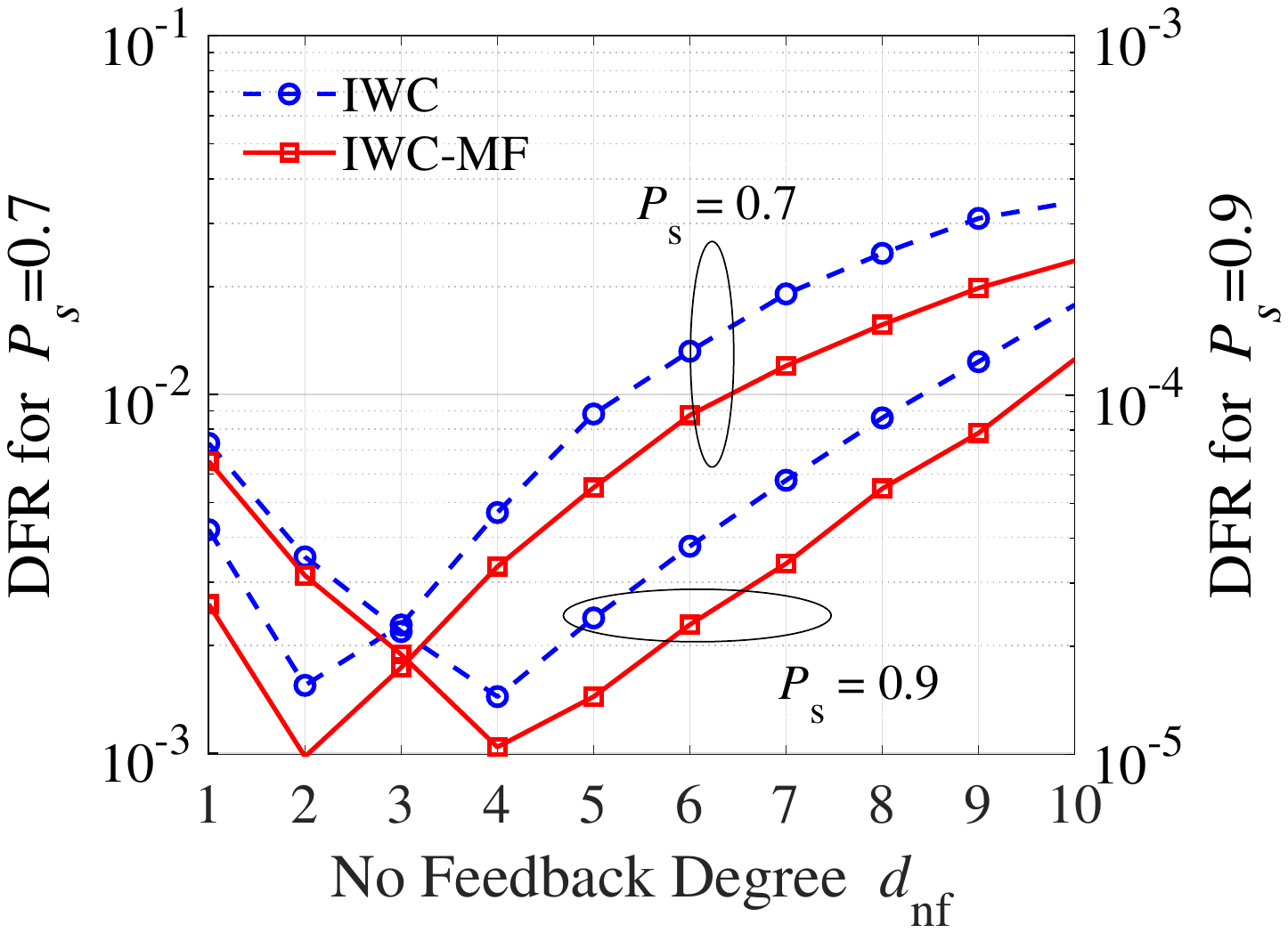}
    \caption{Impact of the no-feedback degree.}
    \label{fig:DFR_vs_NFD}
\end{figure}

Fig.~\ref{fig:DFR_vs_NFD} shows the impact of the no-feedback degree $d_{\mathrm{nf}}$ on the performance of IWC and IWC-MF. We observe that the optimal $d_{\mathrm{nf}}$ depends on the packet-reception probability $P_s$. In general, the optimal $d_{\mathrm{nf}}$ decreases with decreasing $P_s$. Recall that for the destination to be able to recover an information symbol from a coded symbol of degree $d_{\mathrm{nf}}$, it is necessary that exactly one of the $d_{\mathrm{nf}}$ information symbols is missing at the destination. For small $P_s$, more symbols are likely to be missing at the receiver, thus the $d_{\mathrm{nf}}$ that maximizes the probability that the destination is missing exactly one out of $d_{\mathrm{nf}}$ information symbols is smaller when $P_s$ is low. For most scenarios of practical interest that we simulated, we observed the optimal value to be between 2 and 4.        

\begin{figure}
    \centering
    \includegraphics[scale=0.56, bb=40 270 820 550]{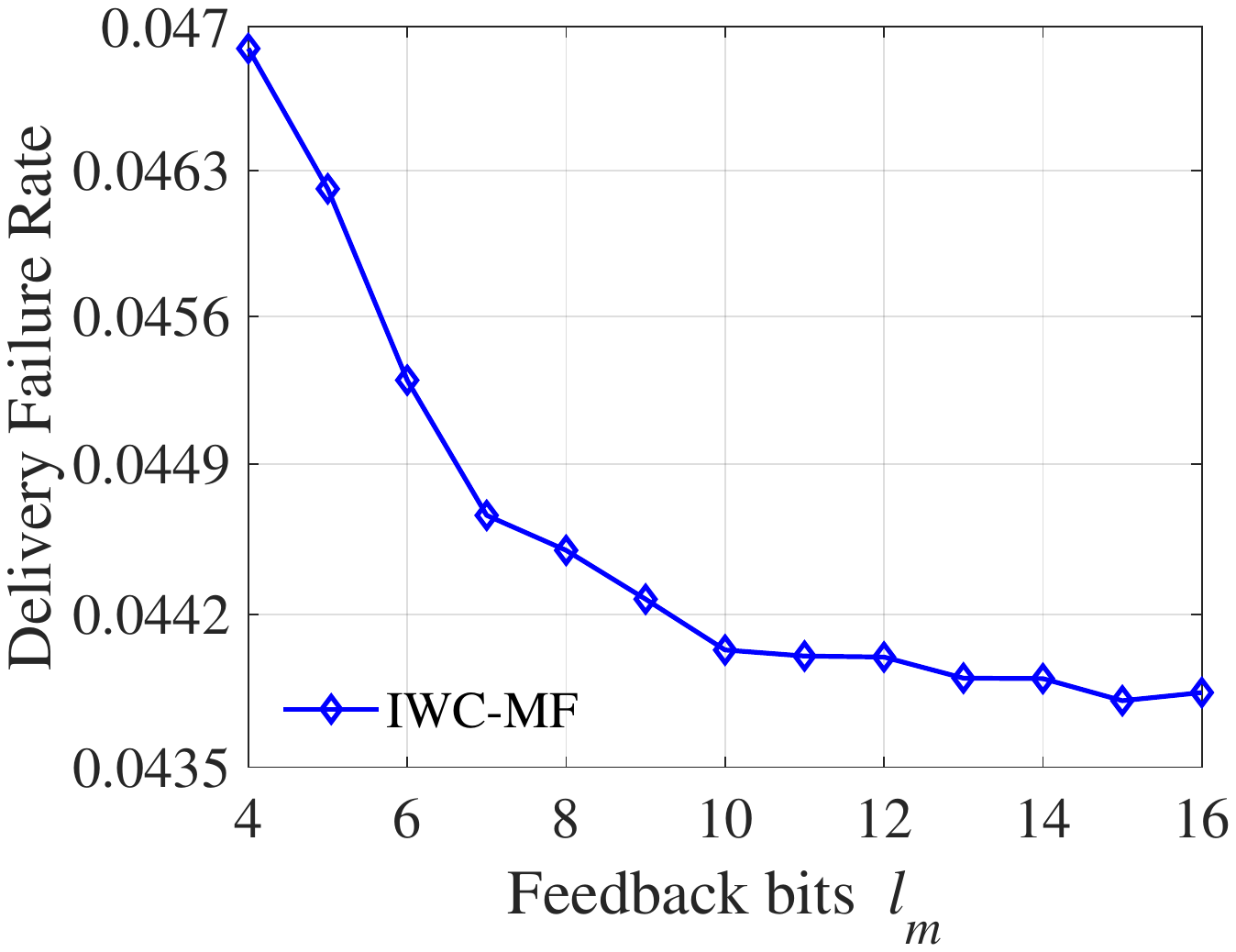}
    \caption{Impact of the number of feedback bits.}
    \label{fig:DFR_vs_lm}
\end{figure}

Fig.~\ref{fig:DFR_vs_lm} shows the performance of IWC-MF as $l_m$, the number of feedback bits that represent the delivery status of the information symbols, is varied. Recall that for IWC, these bits are used to convey the number of unexpired missing symbols. With a delay tolerance of 16, IWC  requires 4 bits to be able to convey any possible number of missing symbols, and further increasing $l_m$ will have no impact on  IWC's performance. IWC-MF, by contrast, requires $l_m \equals 16$ to be able to convey all possible missing patterns. But Fig.~\ref{fig:DFR_vs_lm} shows that the performance penalty incurred by using $l_m \equals 4$ is not large. Also, increasing $l_m$ beyond 10 hardly provides any benefits.      

\begin{figure}
    \centering
    \includegraphics[scale=0.56, bb=40 270 820 550]{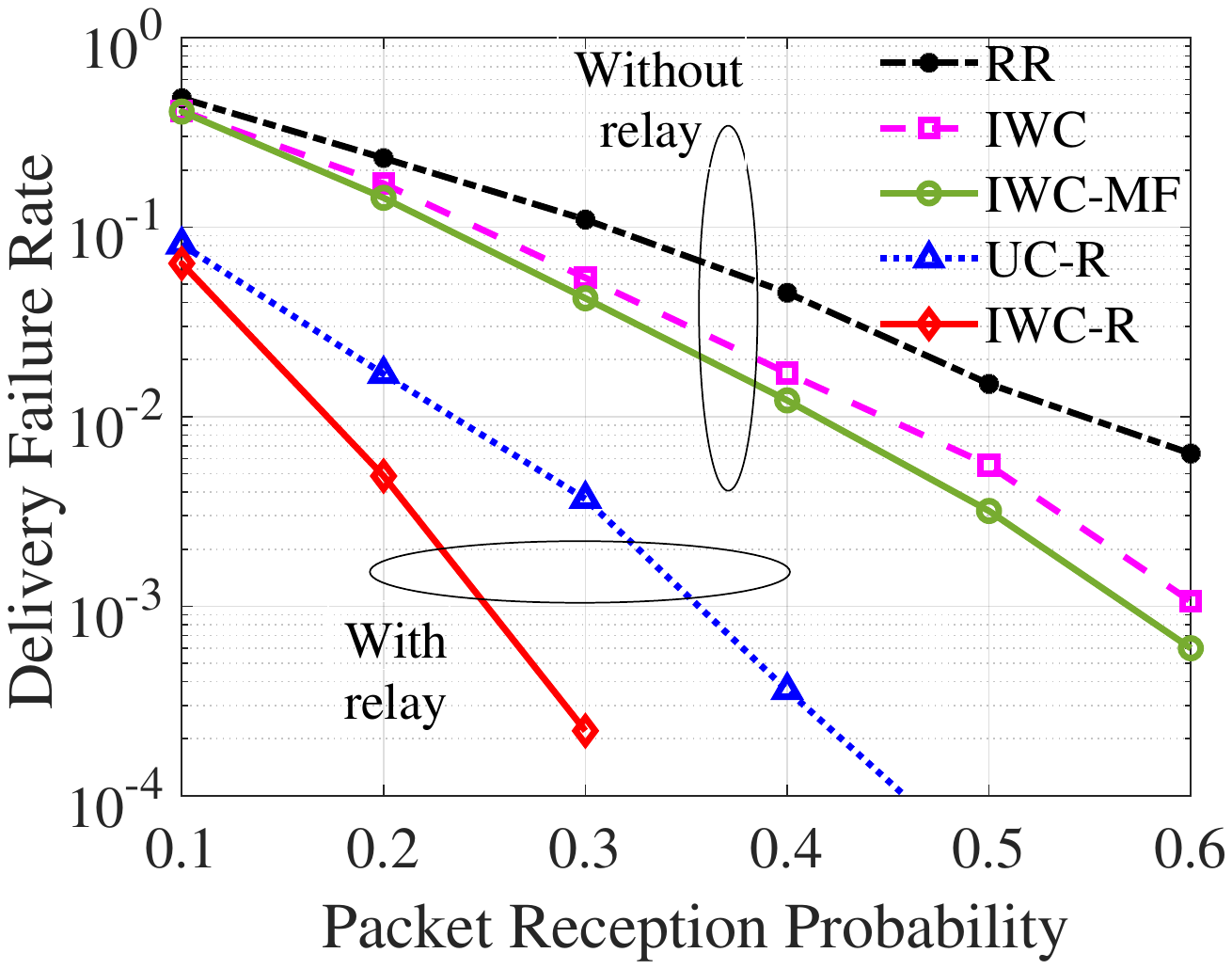}
    \caption{Relay performance on a Bernoulli erasure channel.}
    \label{fig:DFR_Relay_Bernoulli}
\end{figure}

The benefits of relaying are shown in Fig.~\ref{fig:DFR_Relay_Bernoulli} for Bernoulli erasure channels. For IWC-R, the relay threshold is $R_t \equals 2$. As expected, substantially lower DFR is obtained for both coded and uncoded relaying compared to a system without a relay. However, the benefits of the proposed relaying scheme IWC-R is evident, as it provides up to an order of magnitude improvement over UC-R.  Moreover, since $R_t \equals 2$, the relay in IWC-R makes a transmission after every two successful receptions, as opposed to a transmission by UC-R after every successful reception. Thus, the relay spends approximately half the transmission energy  in IWC-R compared to UC-R.  

\begin{figure}
    \centering
    \includegraphics[scale=0.56, bb=40 270 820 550]{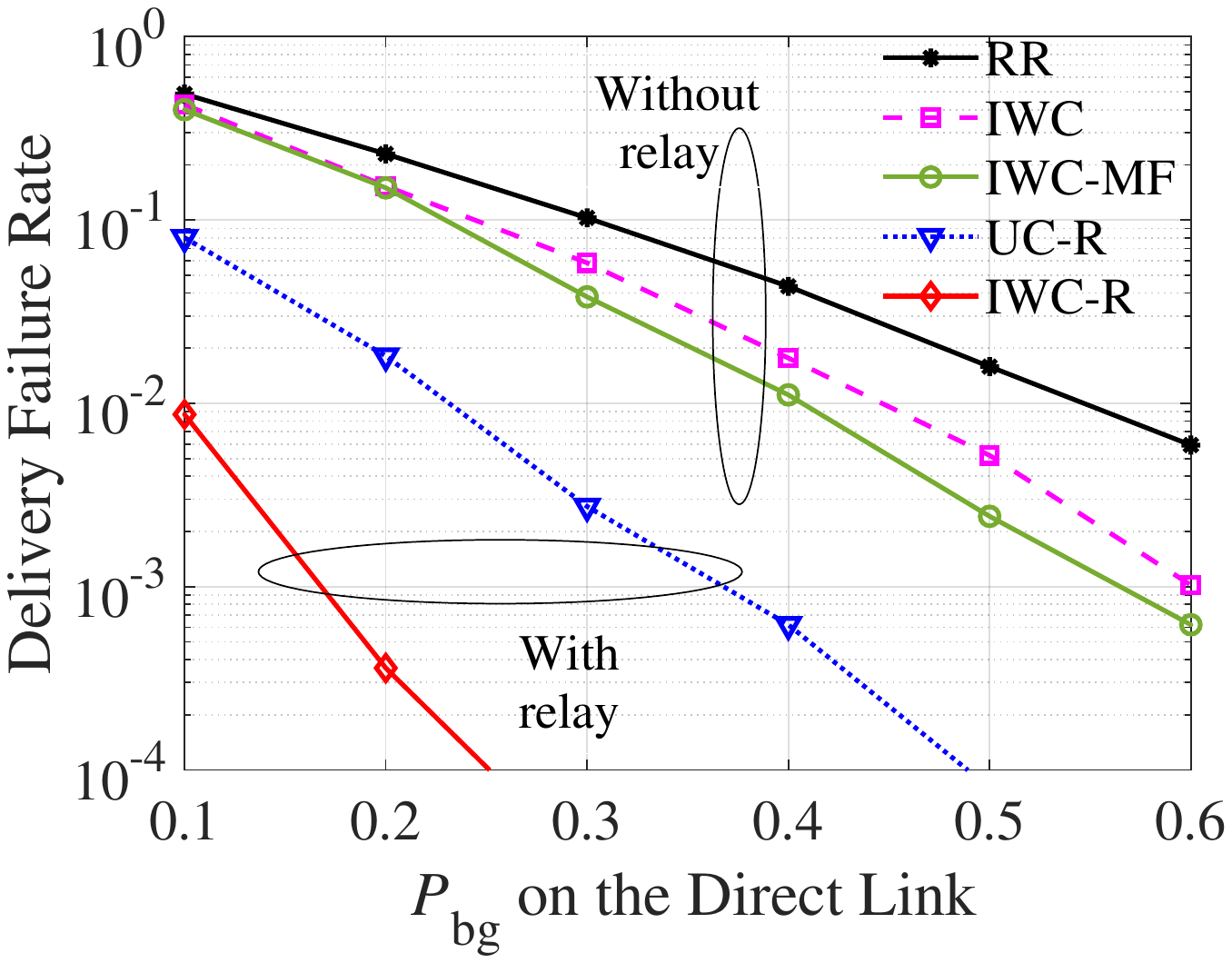}
    \caption{Performance comparison for a Gilbert-Elliott channel.}
    \label{fig:DFR_Relay_GE}
\end{figure}

The performance of the schemes on a Gilbert-Elliott erasure channel are shown in Fig.~\ref{fig:DFR_Relay_GE}. The transition probability $P_\mathrm{gb}$ is fixed at 0.25, while  $P_\mathrm{bg}$ is varied (shown on the x-axis). For IWC-R,  $R_t \!=\! 5$. The performance trends  are quite similar to those for the Bernoulli erasure channel, with the proposed IWC and IWC-MF schemes outperforming RR by up to an order of magnitude, and the proposed IWC-R outperforming all other schemes by up to several orders of magnitude.

\begin{figure}
    \centering
    \includegraphics[scale=0.56, bb=40 270 820 550]{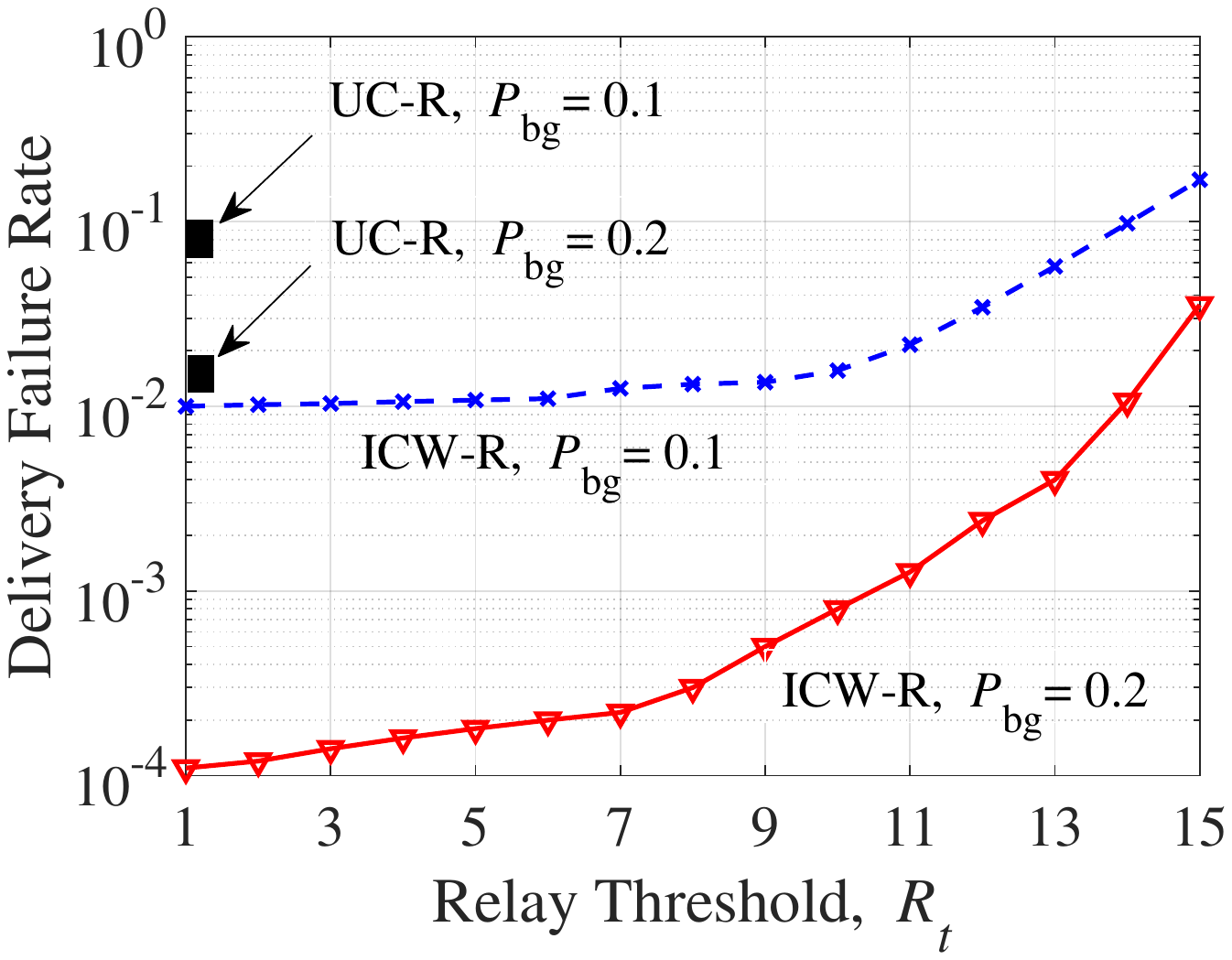}
    \caption{Impact of the relay threshold.}
    \label{fig:DFR_vs_Rt}
\end{figure}

Fig.~\ref{fig:DFR_vs_Rt} shows the impact of the relay threshold $R_t$ on the DFR of IWC-R for two Gilbert-Elliott channels. For each channel, $P_{\mathrm{gb}} \equals 0.25$. We see that a lower $R_t$ provides better delivery performance.  The rectangular markers show the DFR for UC-R for the given $P_{\mathrm{bg}}$. It can be seen that for any $R_t \lthan 14$, ICW-R provides lower DFR than UC-R for either value of $P_{\mathrm{bg}}$.

\begin{figure}
    \centering
    \includegraphics[scale=0.56, bb=40 260 820 550]{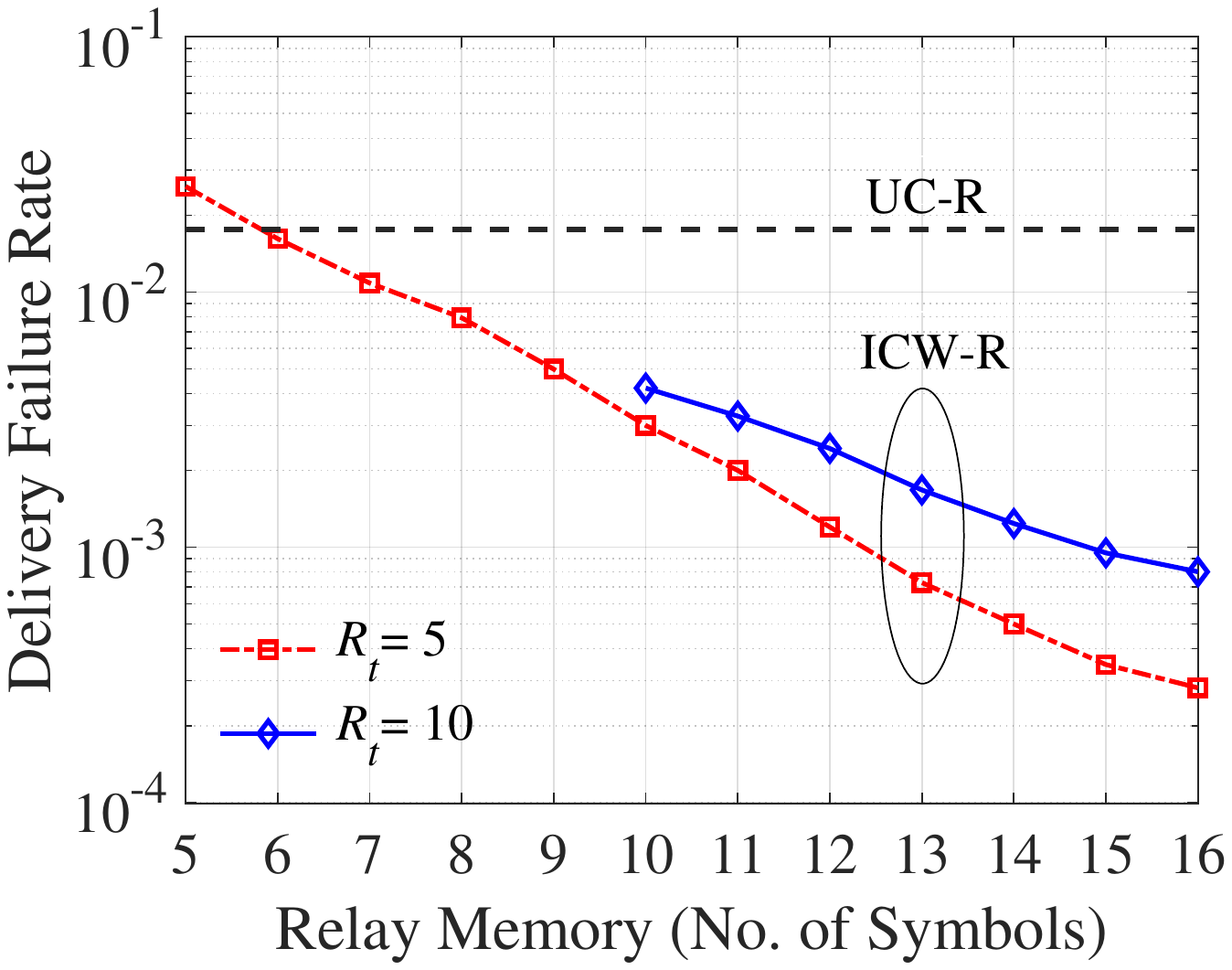}
    \caption{Impact of relay memory.}
    \label{fig:DFR_vs_Rm}
\end{figure}

Note that the maximum relay memory required by \mbox{ICW-R} is equivalent to the number of information symbols that can fit in a time window equal to the application's delay tolerance. In our simulation model, this is equal to $\delta \equals 16$ symbols. However, ICW-R outperforms UC-R even if the relay cannot accommodate 16 symbols at a time. This is illustrated in Fig.~\ref{fig:DFR_vs_Rm}, where we plot the DFR of the schemes as a function of the number of symbols the relay can store. Of course, \mbox{UC-R's} performance is independent of the relay memory since it forwards every information symbols instantaneously. For ICW-R, we consider two relay thresholds, $R_t \equals 5$ and  $R_t \equals 10$, and vary the relay memory from  $R_t$ to $\delta$. We observe significant reductions in the DFR relative to UC-R even when relay memory is less than the maximum required value.

We now consider a practical use case involving energy-monitoring in a smart-factory environment. In particular, we focus on the wireless transfer of energy-consumption data from a laser-cutting tool in a factory site to a monitoring station. The energy-consumption dataset,  made available by FIWARE Lab~\cite{dataset}, was collected as part of the EU FINESCE project at a trial site in the Aachen/Cologne region in Germany, and it provides the energy-consumption information every minute. We consider a scenario in which the information is transmitted using the LoRa communication technology~\cite{Sem13} to a gateway 250 m away from the laser cutting tool. The gateway forwards the received data to the monitoring station which tracks the individual and cumulative energy consumption across the factory. We assume that there are other LoRa modules also installed throughout the factory, each transmitting certain measured parameters. We simulate this scenario using an extended version of the LoRasim simulator~\cite{LoRaSim}. For simulation purposes, the LoRa devices are modeled to be uniformly distributed over a circular area of 300 m radius, with the gateway at the center. The wireless links are subject to independent Rayleigh fading, and have a path loss exponent of 3. The LoRa devices employ spreading factor 7, uses a bandwidth of 125 kHz, and chooses one of 3 non-overlapping channels at random for every transmission. Each measurement is sent in its  32-bit floating point representation (resulting in 4 bytes per symbol) and transmitted as the payload of a LoRa frame (packet). The LoRa devices operate in the confirmed uplink mode, which requires the gateway to acknowledge received frames by sending a feedback frame. Both the feedback  and uplink frames may be lost due to fading or interference from other transmitters. We consider  a delay tolerance of 15 minutes, so that accurate estimates of the average energy consumption in the factory over the past 15 minutes is ideally available at all times. The performance of the coding schemes is shown in Table~\ref{tab:lora_use_case}. We observe that an increase in the number of sensor nodes in the network leads to higher failure rates for all schemes as a result of increased interference; however, IWC and IWC-MF outperforms RR in each case.  

\begin{table}    \caption{Delivery failure rates in a LoRa-based energy monitoring use case with a total of $n$ transmitters (for $b \equals 3)$.}
    \centering
    \begin{tabular}{c|c|c }
    Scheme & $n \equals 50$ & $n \equals 100$ \\
    \hline
        RR & 0.021 & 0.100 \\
        IWC  &  0.012 & 0.071\\
        IWC-MF  & 0.003 & 0.057\\
    \end{tabular}
    \label{tab:lora_use_case}
\end{table}

We now examine the increase in the duty cycle due to the insertion of extra symbols in a frame in the context of the aforementioned use case. The duty cycle is given by
\begin{align}
    \mathrm{DC} = \frac{l_f(b)}{T_p},
\end{align}
where $l_f(b)$ is the length of a frame carrying $b$ symbols and $T_p$ is the interval between two consecutive frames. For LoRa,
\begin{align} \nonumber
    l_f(b) &= \bigg[(n_{\mathrm{pr}} + 4.25) +  8 + \\
        & \max\left\{ \left\lceil \frac{2bm - \SF \minus 5h + 11}{\SF \minus 2q} \right\rceil (c + 4), 0\right\}\bigg]\,\frac{2^{\SF}}{w}, 
\end{align}
where  $\SF$ is the spreading factor, $m$ is the number of bytes per symbol, $w$ is the transmission bandwidth, $n_{\mathrm{pr}}$ is the number of preamble symbols, $h$ is 0 or 1 depending on whether an optional header is included or not, $q$ is 1 when low data rate optimization is enabled and 0 otherwise, and $c$ takes integer values between 1 and 4 depending on the channel code~\cite{Sem13}. 
In general, the duty cycle is related to the coding rate (i.e., the ratio of the number of information symbols in a frame to the total number of information plus coded symbols), as more coded symbols lead to longer transmissions. Note however that finding the coding rate for the proposed schemes is not straightforward since out of the $b \minus 1$ additional symbols in the frame, not all may be coded symbols. To consider the worst-case rate, we define the \textit{minimum coding rate} (MCR) to be $1/b$. Table~\ref{tab:duty_cycle} shows the duty cycle, minimum coding rate, and the DFR for the use case described above. We observe that despite steep reductions in the coding rate, the duty-cycle increase is rather small. This is a consequence of the fact that a transmitted frame has a fixed-size header whose length is often substantial relative to the payload in a typical sensor network application. Thus, an  $m$-folds increase in the payload size leads to much less than an $m$-folds increase in the frame length and the resulting duty cycle.                 

\begin{table}[]    \caption{DFR for IWC and IWC-MF vis-\'{a}-vis duty cycle requirements. (Frame lengths given in milliseconds.)}
    \centering
    \begin{tabular}{c|c|c|c|l|c}
         $b$ & $l_f(b)$ & MCR & DC & {IWC}  & IWC-MF \\ \hline
         1 & 30.976 & 1.00 & 0.052\% & 0.601 & 0.601 \\
         2 & 36.096 & 0.50 & 0.060\% & 0.265  &  0.264 \\
         3 & 41.216 & 0.33 & 0.069\%  &  0.104  &  0.0632\\
         4 & 51.456 & 0.25 & 0.086\%  & 0.0443 &   0.0157 \\
         5 & 56.576 & 0.20 & 0.094\%  &  0.016 & 0.008\\
         6 & 61.696 & 0.17 & 0.103\%  &  0.008 &  0.006\\
    \end{tabular}
    \label{tab:duty_cycle}
\end{table}

We finally remark on the selective coding (SC) approach from~\cite{BSR20}, introduced as an extension of WC. It is identical to WC, except that the feedback has an ACK/NACK bit for the recent packet. This bit allows the source to know more about the delivery status of the information symbols and exclude some symbols from the coding set while forming coded messages. The enhancements we proposed -- namely the simplified degree selection following a feedback, non-random no-feedback degree, and feedback modification to convey the individual delivery status of the symbols instead of the total number of missing symbols -- are directly applicable to SC in the same way they were applied to WC. Although space constraints prevented us from including numerical results on modifications to SC, our simulations showed performance improvements relative to SC that were similar to the improvements reported here relative to WC.              

\section{Conclusion}
\label{sec:conclusion}
We proposed techniques to recover lost data in the uplink of delay-constrained IoT networks in which the sender sporadically receives feedback from the receiver. Through a combination of uncoded retransmissions and coded transmissions, our proposed techniques provide orders-of-magnitude reductions in the delivery-failure rate compared to uncoded repetition redundancy and a state-of-the-art coded transmission scheme from prior literature. The price to be paid in terms of computational complexity is kept low by producing coded symbols only when necessary and by employing our simple degree-computation mechanism. We extended our approach to scenarios in which a relay node overhears the communications between the source and destination and makes coding and forwarding decisions based on overheard feedback, providing large improvements relative to uncoded relaying.

\begin{appendix}
We first note that our goal is to find the maxima of the function in Eq.~\eqref{eq:original-opt-d}, which is equivalent to the following optimization problem: 

\begin{equation}
	\begin{aligned}
		& \underset{d}{\text{maximize}}
		& & f(d) = \frac{y\cdot \binom{x - y}{d-1}}{\binom{x + 1}{d}} \\
		& \text{subject to}
		& & 1 \leq d \leq i-u-\beta ,\; \\
		&&& d \in  \mathbb{Z}^+ \\
		& \text{where} 
		& & x = i-u-1 \\
		&&& y=\beta-1.
	\end{aligned}
\end{equation}
The function $f(d)$ simplifies to
\begin{align} 
f(d)	 &= d\cdot y\frac{(x-y)\cdot(x-y-1)\dots(x-y-d+2)}{(x+1)\cdot(x-1)\dots(x-d+1)}.
\end{align}
We find the optimal $d$ by maximizing \mbox{$h(d) \equals \ln f(d)$}, that is
\begin{align}  
	f(d) 
	&= \ln d + \ln y + \sum_{k=0}^{d-2}{\ln(x-y-k)} - \sum_{k=-1}^{d-1}{\ln(x-k)}.    
\end{align}
The derivate for the above function can be found using the discrete derivative equation, $y'[n] = y[n] - y[n-1]$. The critical point $d^*$ would then be the null point of the derivative
\begin{equation}
	h'(d) = \ln\left({\frac{d}{d-1}}\right) + \ln(x-y-d+2) - \ln(x-d+1).    
\end{equation}
It is easily found that
\begin{align}  \label{eq:opt-d}
	d^*&= \frac{x+1}{y} = \frac{i-u}{\beta - 1}.
\end{align}
It can also be shown that $h''(d) \lthan 0$ for all values of $d$. Thus, $d^*$ is the only optimum value. There are, however, two issues that must be taken care of. First, it may so happen that the critical point lies outside $[1, i \minus u \minus \beta]$; in that case, we use $d^* \equals i \minus u \minus \beta$ as the optimal degree. Secondly, the optimal $d^*$ given by~\eqref{eq:opt-d} may not be an integer. In that case, we take the floor of the value given by~\eqref{eq:opt-d}. Consequently, our final expression for the degree to be used for the packet following a feedback reception is
\begin{equation} \label{eq:degree_oneshot*}
	d^{*} = \min\left\{\left\lfloor{\frac{i-u}{\beta - 1}}\right\rfloor, i-u-\beta\right\}.
\end{equation}
\end{appendix}

\end{document}